\documentclass[prd,twocolumn,superscriptaddress,altaffilletter,amssymb,showpacs,nofootinbib]{revtex4}
\usepackage[dvips]{graphicx}
\usepackage{amsmath}
\newcommand{\be}{\begin{equation}}
\newcommand{\ee}{\end{equation}}
\newcommand{\bea}{\begin{eqnarray}}
\newcommand{\eea}{\end{eqnarray}}

\begin{document}

\title{Dynamics of a Self-interacting Scalar Field Trapped in the Braneworld for a Wide Variety of Self-interaction Potentials}

\author{Yoelsy Leyva}\email{yoelsy@uclv.edu.cu}
\affiliation{Departamento de F\'{\i}sica, Universidad Central de Las Villas, 54830 Santa Clara, Cuba.}

\author{Dania Gonz\'alez}\email{dgm@uclv.edu.cu}
\affiliation{Departamento de F\'{\i}sica, Universidad Central de Las
Villas, 54830 Santa Clara, Cuba.}

\author{Tam\'e Gonz\'alez}\email{tame@uclv.edu.cu}
\affiliation{Departamento de F\'{\i}sica, Universidad Central de Las
Villas, 54830 Santa Clara, Cuba.}

\author{Tonatiuh Matos\footnote{Part of the Instituto Avanzado de Cosmolog\'ia (IAC) collaboration http://www.iac.edu.mx/}}\email{tmatos@fis.cinvestav.mx}
\affiliation{Departamento de F{\'\i}sica, Centro de Investigaci\'on y de Estudios Avanzados del IPN,\\A.P. 14-740, 07000 M\'exico D.F., M\'exico.}

\author{Israel Quiros}\email{israel@uclv.edu.cu}
\affiliation{Departamento de F\'{\i}sica, Universidad Central de Las
Villas, 54830 Santa Clara, Cuba.}

\date{\today}
\begin{abstract}

We apply the dynamical systems tools to study the linear dynamics of a self-interacting scalar field trapped in the braneworld, for a wide variety of self-interaction potentials. We focus on Randall-Sundrum (RS) and on Dvali-Gabadadze-Porrati (DGP) braneworld models exclusively. These models are complementary to each other: while the RS brane produces ultra-violet (UV) corrections to general relativity, the DGP braneworld modifies Einstein's theory at large scales, i. e., produces infra-red (IR) modifications of general relativity. This study of the asymptotic properties of both braneworld models, shows -- in the phase space -- the way the dynamics of a scalar field trapped in the brane departs from standard general relativity behaviour.

\end{abstract}

\pacs{04.20.-q, 04.20.Cv, 04.20.Jb, 04.50.Kd, 11.25.-w, 11.25.Wx, 95.36.+x, 98.80.-k, 98.80.Bp, 98.80.Cq, 98.80.Jk}
\maketitle

\section{Introduction}

Randall-Sundrum (RS) braneworld models have an appreciable impact on early universe cosmology, in particular, on the inflationary paradigm. Actually, a distinctive feature of cosmology of a scalar field confined to a RS brane is that the expansion rate of the universe differs at high energy from that predicted by standard general relativity. This is due to a term quadratic in the energy density, that produces enhancing of the friction acting on the scalar field. This means that, in RS braneworld cosmology, inflation is possible for a wider class of potentials than in standard cosmology \cite{hawkins}. Even potentials that are not sufficiently flat from the point of view of the conventional inflationary paradigm can produce successful inflation. At sufficiently low energies (much less than the brane tension), the standard cosmic behavior is recovered prior to primordial nucleosynthesis scale ($T\sim 1\; MeV$) and a natural exit from inflation ensues as the field accelerates down its potential \cite{5}.\footnote{In this scenario, reheating arises naturally and radiation is created through gravitational particle production \cite{6} and/or through curvaton reheating \cite{7}. This last ingredient improves the brane "steep" inflationary picture \cite{8}. Other mechanisms such as preheating, for instance, have also been explored \cite{9}.} Another interesting feature of this scenario is that the inflaton does not necessarily need to decay; it may survive through the present epoch in the cosmic evolution. Therefore, it may also play the role of the quintessence field, which is a necessary ingredient to explain the current acceleration of the expansion of the universe. Such a unified theoretical framework for the description of both inflaton and quintessence with the help of just one single scalar field has been the target of some works (see for instance Refs. \cite{5,10,11,12,gmq}).

Another braneworld model that has received much attention in the last years, is the Dvali-Gabadadze-Porrati (DGP) model. It describes a brane with 4D world-volume, that is embedded into a flat 5D bulk, and allows for infrared (IR)/large scale modifications of gravitational laws. A distinctive ingredient of the model is the induced Einstein-Hilbert action on the brane, that is responsible for the recovery of 4D Einstein gravity at moderate scales, even if the mechanism of this recovery is rather non-trivial \cite{deffayet}. The acceleration of the expansion at late times is explained here as a consequence of the leakage of gravity into the bulk at large (cosmological) scales, so it is just a 5D geometrical effect,unrelated to any kind of mysterious "dark energy". As with many IR modifications of gravity, there are ghosts modes in the spectrum of the theory \cite{kazuya}.\footnotemark\footnotetext{In fact there are ghosts only in one of the branches of the DGP model; the so called "self-accelerating" branch, or self-accelerating cosmological phase \cite{lue}. The Minkowski cosmological phase is free of ghosts.} Nevertheless, studying the dynamics of DGP models continues being a very attractive subject of research. It is due, in part, to the very simple geometrical explanation to the "dark energy problem", and, in part, to the fact that it is one of a very few possible consistent IR modifications of gravity that might be ever found.

Aim of this paper is to extend the study of references \cite{plb2008,isra} -- the investigation of the dynamics of a self-interacting scalar field trapped on a DGP brane, and on a RS braneworld, respectivley -- to include a wide variety of self-interaction potentials beyond the constant and exponential potentials. This goal will allow us to make conclusive arguments in favour of (or against) the claim made in \cite{gmq} about the genericity of unification of the inflaton and of the quintessence in the Randall-Sundrum scenario. We expect a similar result regarding genericity of gravitational screening of the potential energy of the scalar field within the DGP brane context. 

As in references \cite{plb2008,isra} here we make use of the dynamical systems tools to retrieve useful information on the asymptotic properties of the models under study \cite{coley}. In order to be able to analyze self-interaction potentials beyond the exponential one we will rely on a method proposed recently in the reference \cite{fang}. 

The organization of the paper is as it follows. In section II we provide important details about the Randall-Sundrum model. These include the field equations, the phase space variables chosen, and the mathematical definition of the phase space itself. The same features, this time for the Dvali-Gabadadze-Porrati braneworld model, are given in section III. The results of the study of the corresponding critical points and their stability properties are shown in section IV. The section V is aimed at the physical discussion of the above results, while the conclusions are given in section VI. Through the paper we use natural units ($8\pi G=8\pi/m_{Pl}^2=\hbar=c=1$).

\section{The Randall-Sundrum Model}

We will be concerned here with the dynamics of a self-interacting scalar field with an arbitrary self-interaction potential, that is trapped in a Randall-Sundrum brane of type 2 (RS2). The field equations -- using the Friedmann-Robertson-Walker (FRW) metric -- are the following:

\bea &&3H^2=\rho_T(1+\frac{\rho_T}{2\lambda}),\label{friedmann'}\\
&&2\dot H=-(1+\frac{\rho_T}{\lambda})(\dot\phi^2+\gamma\rho_m),\label{raycha}\\
&&\dot\rho_m=-3\gamma H\rho_m,\;\;\ddot\phi+\partial_\phi V=-3H\dot\phi.\label{continuity'}
\eea where $\lambda$ is the brane tension, $\gamma$ is the barotropic index of the background fluid, $\rho_{T}=\rho_{\phi}+\rho_{m}$ and $V$ is the scalar field self-interaction potential.

Following \cite{isra} we introduce the following dimensionless phase space variables in order to build an autonomous system out of the above system of cosmological equations:

\be  x\equiv\frac{\dot\phi}{\sqrt{6}H},\;y\equiv\frac{\sqrt{V}}{\sqrt{3}H},\;z\equiv\frac{\rho_T}{3H^2}.\label{var'}\ee After this choice of phase space variables we can write the following autonomous system of ordinary differential equations (ODE):

\bea&&x'=-\sqrt\frac{3}{2}y^2 (\partial_{\phi} \ln V) -3x+\nonumber\\
&&\;\;\;\;\;\;\;\;\;\;\;\;\;\;\frac{3}{2}x[2x^2+\gamma(1-x^2-y^2)],\label{eqx}\\
&&y'=\sqrt\frac{3}{2}(\partial_{\phi} \ln V) xy+\nonumber\\&&
\;\;\;\;\;\;\;\;\;\;\;\;\;\;\frac{3}{2}y[2x^2+\gamma(1-x^2-y^2)],\label{eqy}\\
&&z'=3(1-z)(2x^2+\gamma(z-x^2-y^2)),\label{eqz}\eea where the comma denotes derivative with respect to the new time variable ($\tau\equiv\ln a$). Notice that, after the above choice of variables one can realize that

\be \frac{\rho_T}{\lambda}=\frac{2(1-z)}{z},\;\Rightarrow\;0<z\leq 1.\ee This means that the four-dimensional (4D)/low-energy limit of the Randall-Sundrum cosmological equations -- corresponding to the formal limit $\lambda\rightarrow\infty$ -- can be associated with the value $z=1$. The high-energy limit $\lambda\rightarrow 0$, on the contrary, corresponds to $z\rightarrow 0$. The critical points associated with $z=0$, if any, have to be analyzed carefully. Actually, in connection with the classical character of the underlying theory of gravity, the physical meaning of these points in phase space has to be taken with caution due to the high energies associated with them. 

As long as one considers just constant and exponential self-interaction potentials ($\partial_\phi V=0$ and $\partial_\phi V=const$ respectively), the equations (\ref{eqx}-\ref{eqz}) form a closed autonomous system of ordinary differential equations. However, if one wants to go further to consider a wider class of self-interaction potentials beyond the exponential one, the system of ordinary differential equations (\ref{eqx}-\ref{eqz}) is not a closed system of equations any more, since, in general, $\partial_\phi V$ is a function of the scalar field $\phi$. A way out of this difficulty can be based on the method developed in \cite{fang}.

In order to be able to consider arbitrary self-interaction potentials one needs to consider one more variable $s$, that is related with the derivative of the self-interaction potential through $s\equiv-\partial_\phi V/V=-\partial_\phi\ln V$. Hence, an extra equation

\be s'=-\sqrt{6} x s^2 (\Gamma-1),\label{nueva}\ee  has to be added to the above autonomous system of equations. The quantity $\Gamma\equiv V\partial_\phi^2 V/(\partial_\phi V)^2$ in equation (\ref{nueva}) is, in general, a function of $\phi$. The idea behind the method in \cite{fang} is that $\Gamma$ can be written as a function of the variable $s\in\Re^+$, and, perhaps, of several constant parameters. Indeed, for a wide class of potentials the above requirement: $\Gamma=\Gamma(s)$, is fulfilled, see Table \ref{tab1}. 

As in \cite{fang} we introduce a new function $f(s)=\Gamma(s)-1$ so that equation (\ref{nueva}) can be written in the more compact form:

\be s'=-\sqrt{6} x s^2 f(s).\label{nueva'}\ee Equations (\ref{eqx}-\ref{eqz},\ref{nueva'}) form a four-dimensional closed autonomous system of ordinary differential equations, that can be safely studied with the help of the standard dynamical systems tools \cite{coley}. 

The phase space for the autonomous dynamical system driven by the evolution equations (\ref{eqx}-\ref{eqz},\ref{nueva'}) can be defined as it follows: 

\bea &&\Psi=\{(x,y,z):
0\leq x^2+y^2\leq z,\nonumber\\&&-1\leq x\leq
1,0\leq y,0<z\leq 1\}\times\{s\in\Re^+\}.\label{phasespace}\eea

\begin{table*}[tbp]\caption[crit]{Explicit form of the function $\Gamma$ for several quintessential potentials.}
\begin{tabular}{@{\hspace{4pt}}c@{\hspace{14pt}}c@{\hspace{14pt}}c}
\hline\hline\\[-0.3cm]
$\Gamma(s)$ & Potential & Reference \\[0.1cm]
\hline\\[-0.2cm]
$1+\frac{1}{\alpha}-\frac{\alpha\lambda^2}{s^2}$ & $V=V_0
\sinh^{-\alpha}(\lambda\phi)$ & \cite{sahni1} \\[0.2cm]
$\frac{s^2+2m s^2+8m\lambda+s\sqrt{s^2+8m\lambda}}{2 m s^2}$ & $V=V_0
\frac{\exp[\lambda\phi^2]}{\phi^m}$ & \cite{brax} \\[0.2cm]
$1-\frac{1}{2\alpha}+\frac{\alpha\lambda^2}{2s^2}$ & $V=V_0\left[\cosh(\lambda\phi)-1\right]^p$ & \cite{sahni2} \\[0.2cm]
$-\frac{\kappa[\kappa\alpha\beta+s(\alpha+\beta)]}{s^2}$ & $V=V_0\left[\exp(\alpha\kappa\phi)+\exp(\beta\kappa\phi)\right]$ & \cite{barreiro} \\[0.2cm]
 $-\frac{\lambda}{s}$ & $V=V_0\exp(-\lambda\phi)+\Lambda$ & \cite{cardenas} \\[0.2cm]
\hline \hline
\end{tabular}\label{tab1}
\end{table*}

\begin{table*}\caption[crit]{Properties of the critical points for the autonomous system
(\ref{eqx}-\ref{eqz},\ref{nueva'}).}
\begin{tabular}{@{\hspace{4pt}}c@{\hspace{14pt}}c@{\hspace{14pt}}c@{\hspace{14pt}}
c@{\hspace{14pt}}c@{\hspace{14pt}}c@{\hspace{14pt}}c@{\hspace{14pt}}c@{\hspace{14pt}}
c@{\hspace{14pt}}c@{\hspace{14pt}}c@{\hspace{14pt}}c}
\hline\hline\\[-0.3cm]
$P_i$ &$x$&$y$&$z$&$s$&Existence&$\Omega_{\phi}$&$w_\phi$&$q$ \\[0.1cm]\hline\\[-0.2cm]
$P_1$& $0$&$0$&$1$&$s$ & Always&$0$&undefined& $-1+\frac{3\gamma}{2}$\\[0.2cm]
$P_2$& $0$&$y\in]0,1]$&$y^2$&$0$& "&$y^2$&$-1$&$-1$ \\[0.2cm]
$P_3$& $0$&$1$&$1$&$0$ &  "&$1$&$-1$&$-1$ \\[0.2cm]
$P_4$& $1$&$0$ &$1$&$0$& " &$1$&$1$&$2$\\[0.2cm]
$P_5\pm$& $\mp1$&$0$ &$1$&$s_{\ast}$& " &$1$&$1$&$2$ \\[0.2cm]
$P_6$&
$\frac{s_{\ast}}{\sqrt6}$&$\sqrt\frac{6-(s_{\ast})^2}{6}$&$1$&
$s_{\ast}$ &$s_{\ast}^{2}\leq 6$&$1$&$\frac{(s_{\ast})^2-3}{3}$&$\frac{1}{2}(-2+s_{\ast}^{2})$ \\[0.2cm]
$P_7$& $\frac{\sqrt{6}\gamma}{2
s_{\ast}}$&$\sqrt{\frac{3\gamma(2-\gamma)}{2 s_{\ast}^{2}}}$&$1$&
$s_{\ast}$&$s_{\ast}^{2}\geq 3\gamma$&$\frac{3\gamma}{(s_{\ast})^2}$ &$\gamma-1$&$-1+\frac{3\gamma}{2}$\\[0.2cm]
\hline \hline\\[-0.3cm]
\end{tabular}\label{tab1'}
\end{table*}

\begin{table*}\caption[eigenv]{Eigenvalues for the critical points in table \ref{tab1'}.
 $A\equiv\sqrt{(2-\gamma)(24\gamma^{2}-s_{\ast}^{2}(9\gamma-2))}$}
\begin{tabular}{@{\hspace{4pt}}c@{\hspace{14pt}}c@{\hspace{14pt}}c@{\hspace{14pt}}
c@{\hspace{14pt}}c@{\hspace{14pt}}c@{\hspace{14pt}}c@{\hspace{14pt}}c}
\hline\hline\\[-0.3cm]
$P_i$ & $\lambda_1$& $\lambda_2$& $\lambda_3$& $\lambda_4$\\[0.1cm]\hline\\[-0.2cm]
$P_1$& $0$&$\frac{3}{2}(-2+\gamma)$&$-3\gamma$&$\frac{3\gamma}{2}$\\[0.2cm]
$P_2$& $-3$&$0$&$0$&$-3\gamma$\\[0.2cm]
$P_3$& $-3$&$0$&$0$&$-3\gamma$\\[0.2cm]
$P_4$& $-6$&$3$&$0$&$6-3\gamma$\\[0.2cm]
$P_5^\pm$&$-6$&$\pm\sqrt6 s_{\ast}^{2}df$&$3\pm\sqrt\frac{3}{2}s_{\ast}$&$6-3\gamma$\\[0.2cm]
$P_6$& $-s_{\ast}^{2}$&$-s_{\ast}^{2}s_{\ast}df$&$\frac{1}{2}(-6+s_{\ast}^{2})$& $s_{\ast}^{2}-3\gamma$  \\[0.2cm]
$P_7$&$-3\gamma$&$-3\gamma
s_{\ast}df$&$\frac{3}{4}(-2+\gamma)-\frac{3}{4s_{\ast}}A$&$\frac{3}{4}
(-2+\gamma)+\frac{3}{4s_{\ast}}A$\\[0.2cm]
\\[0.4cm]
\hline \hline
\end{tabular}\label{tab2'}
\end{table*}

\section{The Dvali-Gabadadze-Porrati Model}

In this section we will focus our attention in a braneworld model where a self-interacting scalar field is trapped on a DGP brane. In the flat FRW metric, the field equations are the following:

\bea &&Q^2=\frac{1}{3}(\rho_m+\frac{\dot{\phi}^2}{2}+V(\phi)),\label{friedmann}\\
&&\dot\rho_m=-3\gamma H\rho_m,\;\;\ddot\phi+\partial_\phi
V=-3H\dot\phi.\label{continuity}\eea where $\rho_m$ is the energy density of the background barotropic fluid ($\gamma$ is its barotropic index), $V$ its self-interaction potential, $\phi$ is the scalar field trapped in the DGP brane, and 

\be Q^2_\pm\equiv H^2\pm\frac{1}{r_c}H,\label{Qequation}\ee where, as customary, $r_c$ is the crossover scale inherent in the DGP brane model. There are two possible branches of the DGP model corresponding to the two possible choices of the signs in (\ref{Qequation}): "+" is for the Minkowski cosmological phase of DGP model -- that is free of ghosts --, while "-" is for the self-accelerating solution.

Following the reference \cite{plb2008} we define the phase space dimensionless variables:

\be
x\equiv\frac{\dot\phi}{\sqrt{6}Q},\;y\equiv\frac{\sqrt{V}}{\sqrt{3}Q},\;z\equiv\frac{Q}{H}.\label{var}\ee The corresponding autonomous system of equations in the variables $x$, $y$ and $z$, defined above, was used in \cite{plb2008} to study the asymptotic properties of the DGP-quintessence model with constant and exponential potentials, exclusively. The study of other potentials was not considered. 

As in the former section, following a method developed in \cite{fang}, here we extend the analysis of the three-dimensional autonomous system of reference \cite{plb2008}, to four-dimensions, through the addition of the extra-variable $s\equiv -\partial_\phi V/V$ defined in the former section. This will permit us to consider a wider class of self-interaction potentials beyond the exponential one. In consequence, to the system of equations of \cite{plb2008}, we add (\ref{nueva'}), so that we are left with the following autonomous closed system of ordinary differential equations:

\bea &&x'=\sqrt\frac{3}{2}y^2 z s-3x+\nonumber\\&&\;\;\;\;\;\;\;\;\;\;\;\;\;\;\frac{3}{2}x[2x^2+\gamma(1-x^2-y^2)],\label{eqx1}\\
&&y'=-\sqrt\frac{3}{2}x y z s+\nonumber\\&&\;\;\;\;\;\;\;\;\;\;\;\;\;\;\frac{3}{2}y[2x^2+\gamma(z-x^2-y^2)],\label{eqy1}\\
&&z'=\frac{3}{2}z\frac{z^2-1}{z^2+1}[2x^2+\gamma(z-x^2-y^2)],\label{eqz1}\\
&&s'=-\sqrt{6} x s^2 f(s),\label{eqs}\eea where, as before, the comma denotes derivative with respect to the time variable $\tau\equiv\int H dt$, and $f(s)=\Gamma(s)-1$ ($\Gamma\equiv V\partial_\phi^2 V/(\partial_\phi V)^2$). 

After the above choice of phase space variables, the equation (\ref{Qequation}) can be put into the following form: 

\be z^2=1\pm\frac{1}{r_c H}.\label{QequationZ}\ee.

For the Minkowski phase, since $0\leq H\leq\infty$ (we consider just non-contracting universes), then $1\leq z\leq\infty$. The case $-\infty\leq z\leq -1$ corresponds to the time reversal of the later situation. For the self-accelerating phase, $-\infty\leq
z^2\leq 1$, but since we want real valued $z$ only, then $0\leq z^2\leq 1$.\footnotemark\footnotetext{In fact, fitting SN observations requires $H\geq r_c^{-1}$ in order to achieve late time acceleration (see, for instance, reference \cite{kazuya} and
references therein). This means that $z$ has to be real-valued.} As before, the case $-1\leq z\leq 0$ represents time reversal of the case $0\leq z\leq 1$ that will be investigated here. Both branches share the common subset $(x,y,z=1)$, which corresponds to the formal
limit $r_c\rightarrow\infty$ (see equation (\ref{Qequation})), i.e., this represents just the standard four-dimensional behaviour typical of Einstein-Hilbert theory coupled to a self-interacting scalar field.

The phase space for the autonomous system (\ref{eqx1}-\ref{eqs}), for the "+" branch can be defined as: 

\bea \Psi_+=\{(x,y,z):\;0\leq x^2+y^2\leq 1,\nonumber\\
z\in [1,\infty[\}\times\{s\in\Re^+\},\label{PhaseSpace+}\eea while, for the self-accelerating "-" phase, it is given by the non-compact region: 

\bea \Psi_-=\{(x,y,z):\;0\leq x^2+y^2\leq 1,\nonumber\\
z\in ]0,1]\}\times\{s\in\Re^+\}.\label{PhaseSpace-}\eea

Notice that the points belonging in the set $(x,y,0)$ can not be included since, in this case ($z=0\;\Rightarrow\;Q=0$), the variables $x$ and $y$ are undefined. The self-accelerating solution $H=1/r_c$ ($Q_-=0\;\Rightarrow\;z=0$) has been studied in \cite{plb2008}. In that reference the analysis of the critical points of the quintessence model under study was based on a concrete form of the self-interaction potential. Here, as in the former section, we use the approach proposed in \cite{fang} to investigate the critical points of the dynamical system for arbitrary functions $f(s)$, so that, in principle, we are able to study arbitrary self-interaction potentials.

\section{Critical Points of the Autonomous Dynamical System}

In this section we will analyse in detail the critical points of the autonomous systems corresponding to both Randall-Sundrum and Dvali-Gabadadze-Porrari braneworld models, as well as their stability properties.

\subsection{The Randall-Sundrum braneworld}

The critical points of system (\ref{eqx}-\ref{eqz},\ref{nueva'}) are summarized in table \ref{tab1'}. The eigenvalues of the corresponding jacobian matrices are shown in table \ref{tab2'}. In both cases $s_{\ast}$ is the value which makes the function $f(s)$ vanish, i. e., $f(s_{\ast})=0$. In the same way we have chosen $$df\equiv\frac{df(s)}{ds}\mid_{s\ast}.$$

As we see from tables \ref{tab1'} and \ref{tab2'}, the point $P_1$ exist in all cases regardless of the form of the self-interaction potential (arbitrary $s$). Points $P_2-P_4$ are associated with potentials whose first $\phi$-derivative vanishes at some/several point/points (this case includes the constant potential whose $\phi$-derivatives at any order vanish everywhere). Worth noticing that the existence of points $P_5^\pm-P_7$ depends on the concrete form of the potentials (recall that the $s_{\ast}$-s depend on the functional form of $f(s)$). From the table of eigenvalues, notice, besides, that there are four non-hyperbolic critical points/sets of critical points (at least one of the eigenvalues is vanishing): $P_1, P_2, P_3$, and $P_4$. 

We recall that four-dimensional effects are associated with points belonging in the plane $(x,y,z=1)$. For points with $z\neq 1$, five-dimensional effects affect the dynamics of the universe. There is only a set of critical points with $z\neq1$ (represented by $P_2$ in table \ref{tab1'}): $(x,y,z,s)=(0,y,y^{2},0)$. For points in this set, since $x=0$, $z=y^{2}$, while $\rho_{T}=V$, then the Friedmann equation can be written in the form:

\be 3H^{2}=V\left(1+\frac{V}{2\lambda}\right).\label{fr}\ee For values of the potential much larger than the brane tension $V\gg\lambda\;\Rightarrow\;H_{RS}=V/\sqrt{6\lambda}$, so that the early time/high energy expansion rate in the Randall-Sundrum model ($H_{RS}$) gets enhanced with respect to the general relativity rate:

\be \frac{H_{RS}}{H_{GR}}=\sqrt{\frac{V}{2\lambda}}.\label{rate}\ee This is the way brane effects fuel early inflation in the RS model. The fact that this is a critical point in the phase space of the RS model means that, helping inflation to happen is a generic feature of Randall-Sundrum braneworld models. For a further discussion about this see \cite{isra}.

The rest of the critical points of the dynamical system (\ref{eqx}-\ref{eqz},\ref{nueva'}) lie on the plane $(x,y,1)$, so that, only four-dimensional behavior can be associated with them. Non-hyperbolic critical points in the set $P_1=(0,0,1,s)$ correspond to the matter-dominated solution and, as it is seen from table \ref{tab2'}, these are always saddle points in phase space. Critical points $P_2-P_4$ have been exhaustively studied in \cite{plb2008}. 

Points $P_5^\pm$ are saddle critical points. These correspond to the solution dominated by the kinetic energy of the scalar field ($\Omega_{\phi}=1$). This result differs from the one in standard four-dimensional theory, where the kinetic energy-dominated solution can be a past attractor (an unstable source point) for trajectories in the phase space.

The non-hyperbolic point $P_3$ has a two-dimensional stable subspace, which corresponds to a late-time attractor solution ($3H^{2} = V$). 

There are other two critical points that can be associated with late-time attractor solutions: $P_6$ and $P_7$. For values $s_*^2<3\gamma$ ($s_* df>0$), the scalar field-dominated solution (point $P_6$) is the future attractor of the autonomous system (\ref{eqx}-\ref{eqz},\ref{nueva'}). The scaling solution (point $P_7$) is the late-time attractor whenever it exists. 

Due to the fact that there are several non-hyperbolic critical points which can not be consistently studied with the help of the present linear analisys, we choose several concrete examples -- corresponding to different quintessence potentials of cosmological interest -- in order to illustrate, in the phase space, the dynamical behavior of the corresponding RS model, including the neighbourhood of these non-hyperbolic critical points.

\begin{figure}[ht!]
\begin{center}
\includegraphics[width=7cm,height=6cm]{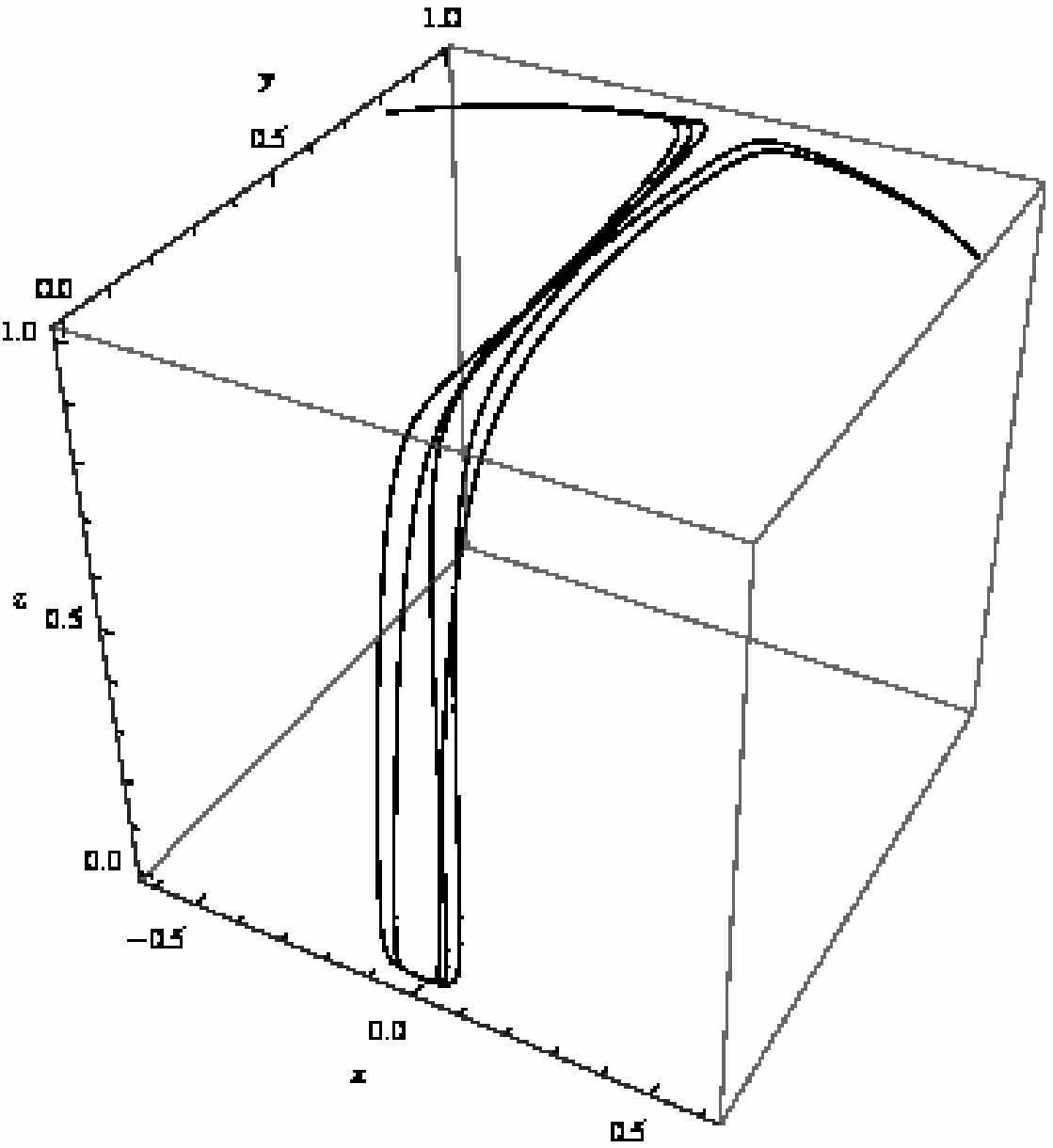}
\includegraphics[width=7cm,height=6cm]{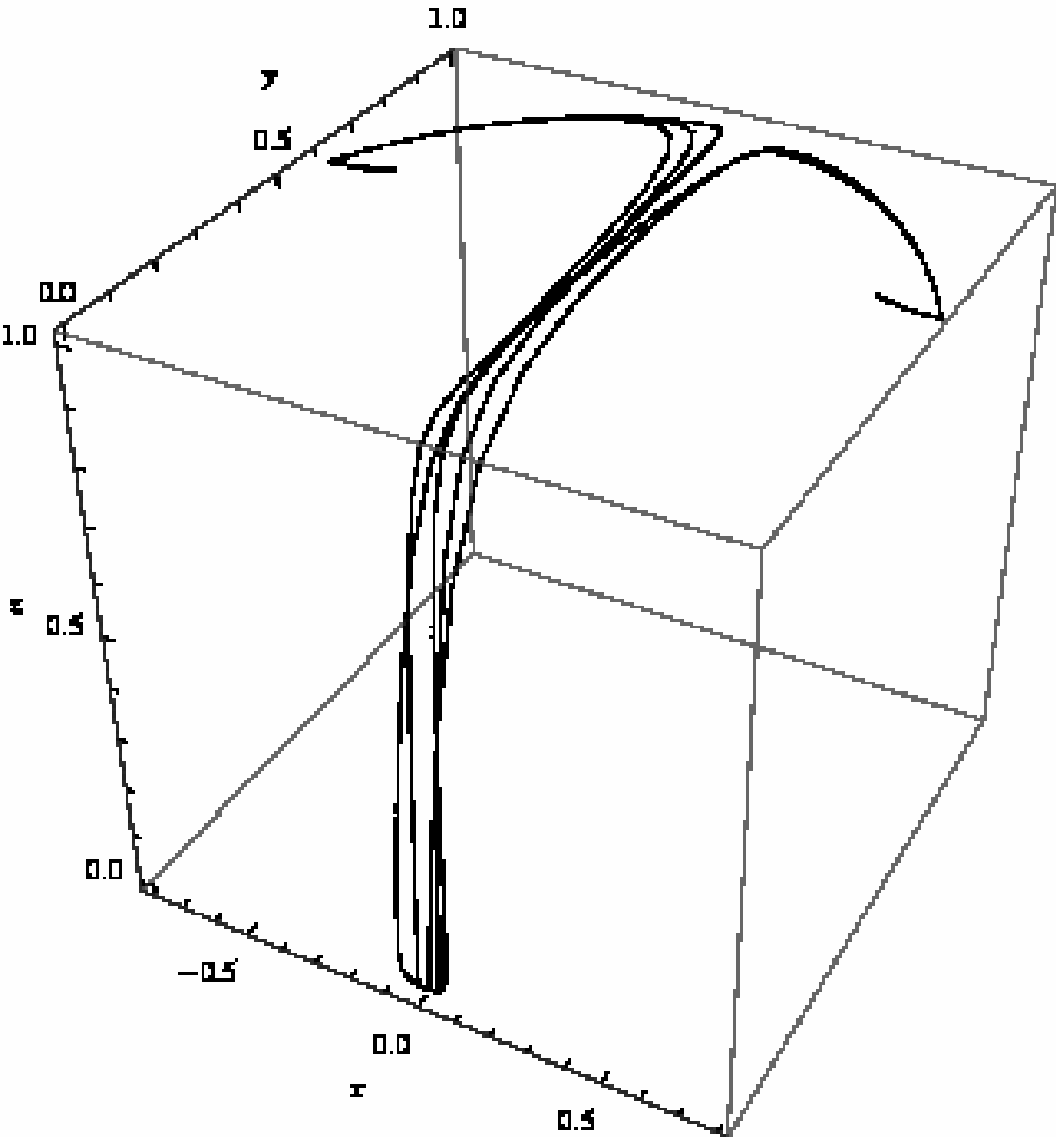}
\includegraphics[width=7cm,height=6cm]{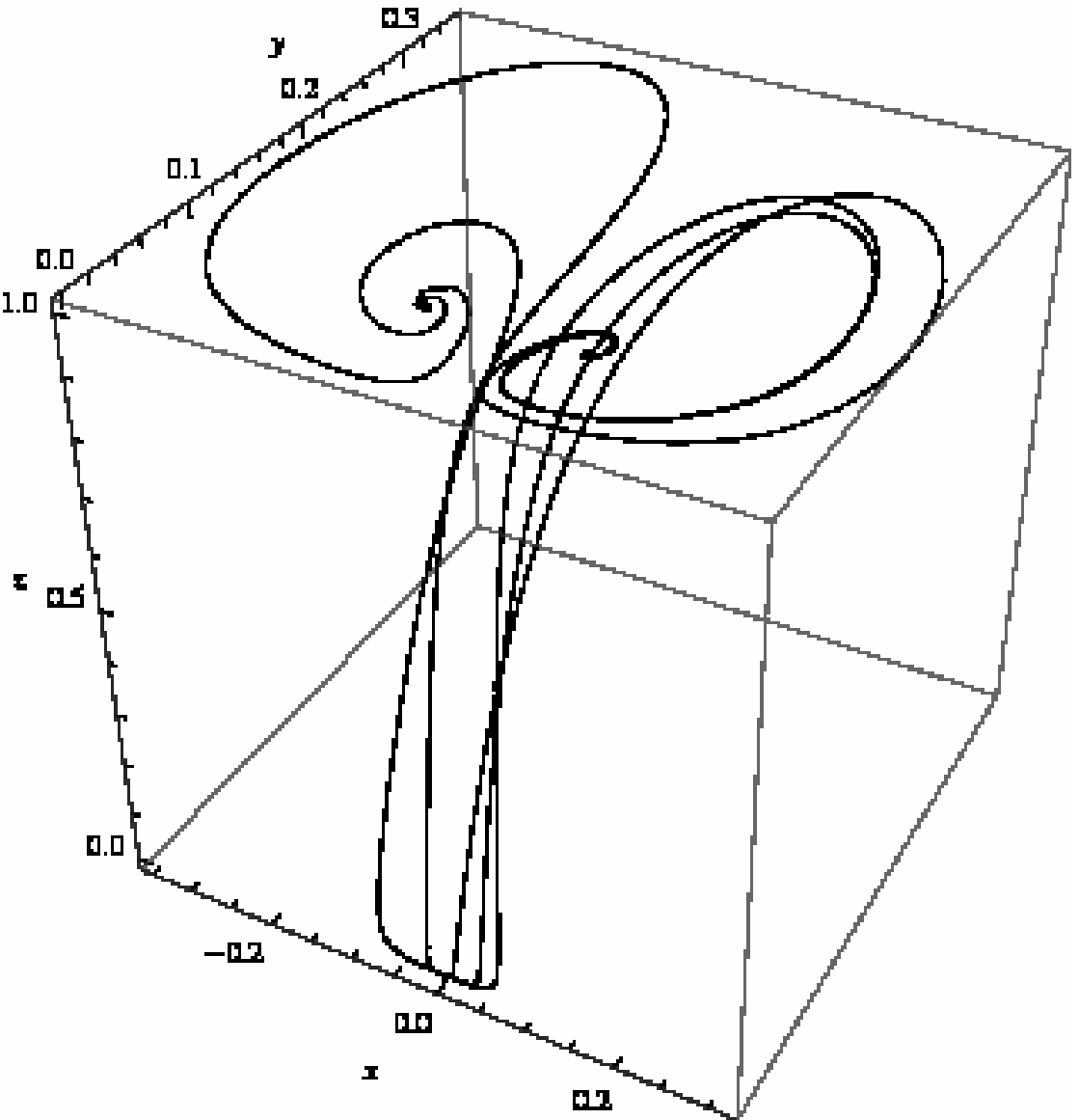}
\vspace{0.3cm}
\caption{Trajectories in phase space $(x,y,z)$ for different sets initial conditions for the potential $V=V_{0}[\sinh(\lambda \phi)]^{-\alpha}$. The free parameters have been chosen to be $(\alpha,\gamma,\lambda)$: $(3,1,0.5)$ - upper panel, $(3,1,0.65)$ - central panel, and $(3,1,5)$ - lower panel. For the first parameter selection the late-time attractor is the scalar field-dominated solution (point $P_6$ in table \ref{tab1'}). For the remaining parameter selections the late-time attractor is the matter-scaling solution (point $P_7$ in table \ref{tab1'}). In the last case this point is a stable spiral.}\label{fig01'}
\end{center}
\end{figure}

\subsubsection{$V=V_{0}[\sinh(\lambda \phi)]^{-\alpha}$}

This potential was studied for the first time in \cite{matos}, where it was shown to be a new cosmological tracker solution for quintessence. For this potential the function $f(s)=1/\alpha-\alpha\lambda^{2}/s^{2}$, while

\be s_{\ast}=\pm \alpha\lambda,\;df=\frac{2\alpha\lambda^{2}}{s_{\ast}^{3}}.
\ee

The scalar field-dominated solution (set of points $P_6$ in tables \ref{tab1'} and \ref{tab2'}) is a late-time attractor whenever $\lambda^{2}\leq 3\gamma/\alpha^2$, and $\alpha>0$. It is consistent with accelerated expansion if $\lambda^{2}<2/\alpha^2$, and $\alpha>0$. Since $s_{\ast}=\pm\alpha\lambda$, $P_7$ is a stable node:

\begin{equation*}
    x_{+}=+\frac{\alpha\lambda}{\sqrt{6}},
    y_{+}=\sqrt{\frac{6-\alpha^{2}\lambda^{2}}{6}}, z_{+}=1, s_{+}=+\alpha\lambda
\end{equation*}
\begin{equation*}
    x_{-}=-\frac{\alpha\lambda}{\sqrt{6}},
    y_{-}=\sqrt{\frac{6-\alpha^{2}\lambda^{2}}{6}}, z_{-}=1, s_{}=-\alpha\lambda
\end{equation*}

Therefore, whenever the condition $3\gamma/\alpha^2<\lambda^{2} < 6/\alpha^2$ ($\alpha>0$) holds true, the matter-scaling solution dominates the late-time evolution of the universe. In fig.\ref{fig01'} this behaviour is illustrated for an arbitrary set of initial conditions. For $\alpha>0$, and $\lambda^{2}>\frac{6}{\alpha^{2}}$, the matter-scaling solution is a stable spiral. It is the late-time attractor in this case.

\begin{equation*}
    x_{+}=+\frac{\sqrt{6}\gamma}{2\lambda\alpha},
    y_{+}=\sqrt{\frac{3\gamma(2-\gamma)}{2\alpha^{2}\lambda^{2}}}, z_{+}=1, s_{+}=+\alpha\lambda
\end{equation*}

\begin{equation*}
    x_{-}=-\frac{\sqrt{6}\gamma}{2\lambda\alpha},
    y_{-}=\sqrt{\frac{3\gamma(2-\gamma)}{2\alpha^{2}\lambda^{2}}}, z_{-}=1, s_{-}=-\alpha\lambda
\end{equation*}

This behavior is clearly shown in fig. \ref{fig01'}. It is seen, in particular, that the trajectories in phase space emerge from the point $S=(x,y,z)=(0,0,0)$ -- the empty Misner-RS universe -- meaning that this is the past attractor of the Randall-Sundrum cosmological model. We want to notice that the points with $z=0$ have been removed from the phase space $\Psi$ since, in general, at $z=0$ the autonomous system of equations (\ref{eqx}-\ref{eqz},\ref{nueva'}) blows up due to our choice of phase space variables. For that reason the point $S$ does not appear in table \ref{tab1'}. See \cite{isra} for further discussion.

\begin{figure}[t!]
\begin{center}
\includegraphics[width=7cm,height=6cm]{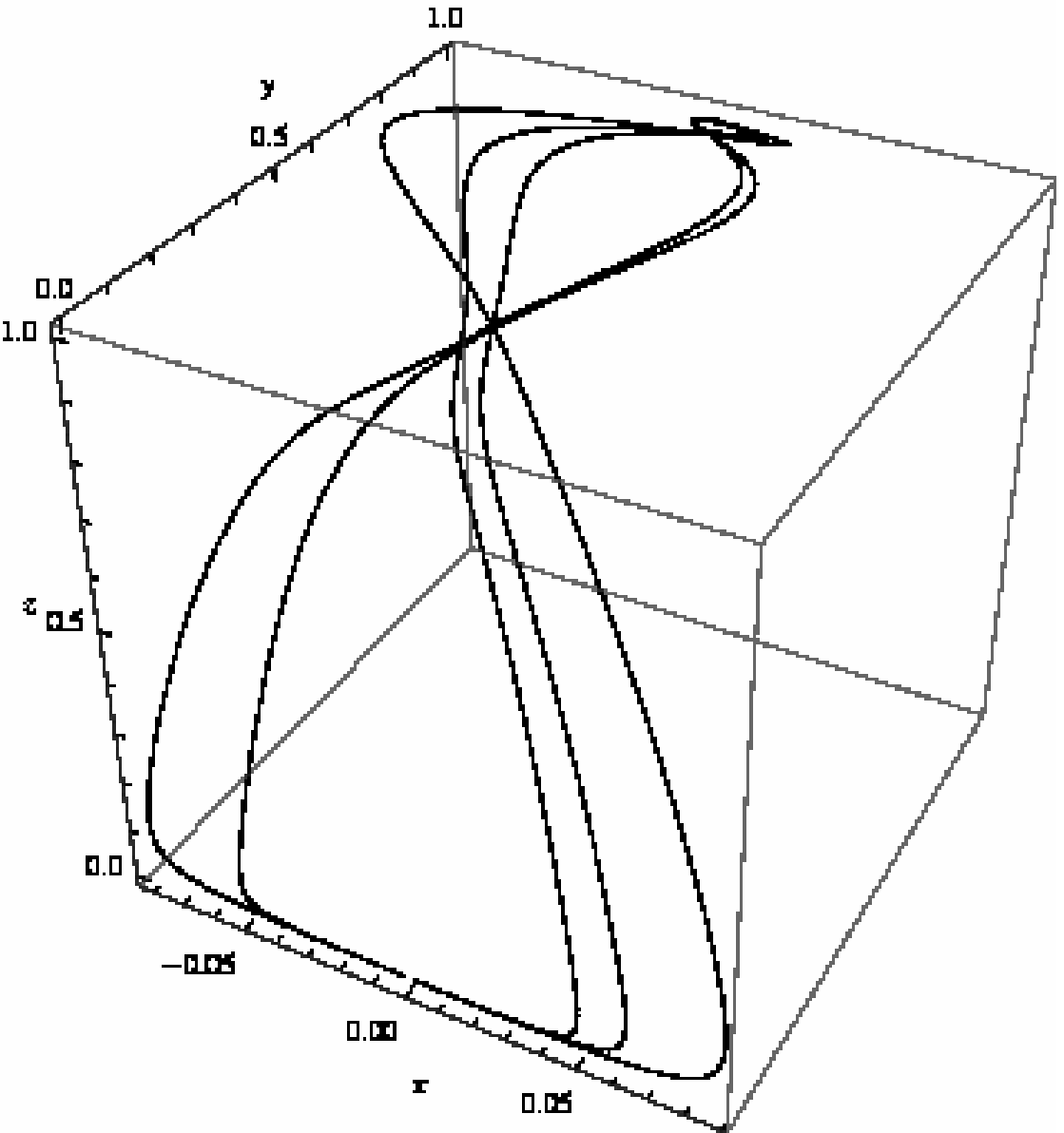}
\includegraphics[width=7cm,height=6cm]{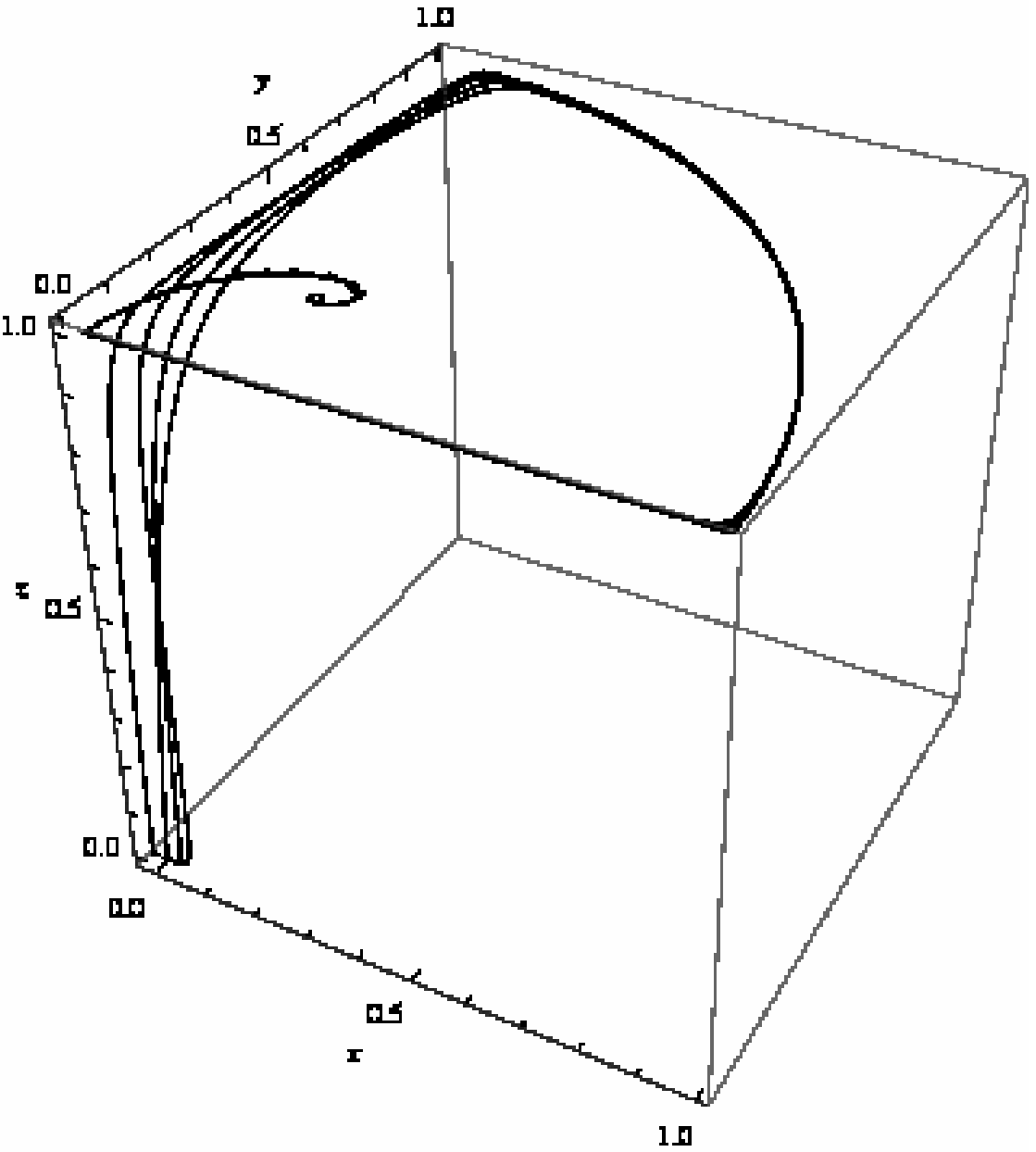}
\vspace{0.3cm}
\caption{Phase trajectories $(x,y,z)$ for given initial data, for the potentials $V=V_{0}[\cosh(\lambda\phi)-1]^{\alpha}$ (upper panel), and $V=\frac{V_{0}}{(\eta+e^{-\alpha\phi})^{\beta}}$ (lower panel). The past attractor in the phase space is the empty, Misner-RS universe (point $S=(x,y,z)=(0,0,0)$ that does not appear in table \ref{tab1'}). In the upper panel we set $\alpha=0.4$, $\lambda=5$ and $\gamma=1$ so that the scalar field-dominated solution (point $P_6$) is the late-time attractor. In the lower panel the matter-scaling solution $P_7$ is a stable spiral ($\alpha=-2.6$, $\beta=2$ and $\gamma=1$). }\label{fig02'}
\end{center}
\end{figure}

\subsubsection{$V=V_{0}[\cosh(\lambda\phi)-1]^{\alpha}$} 

This potential was proposed in \cite{wang} in order to describe both quintessence and a new form of dark matter called frustrated cold dark matter, due to its ability to frustrate gravitational clustering at small scales. Accelerated expansion is obtained for $0<\alpha< \frac{1}{2}$. For this potential

\be f(s)=\frac{1}{2}\left(\frac{\alpha\lambda^{2}}{s^2}-\frac{1}{\alpha}\right),\ee while

\be s_{\ast}=\pm\alpha\lambda,\;df=-\frac{\alpha\lambda^{2}}{s_{\ast}^{3}}.\label{sol_wang}\ee

In order for the critical points $P_6$ and $P_7$ to be late-time attractors, it is necessary that the condition $s_{\ast}df>0$ be fulfilled. For the present potential this condition can be written in the following form:

\begin{equation*} -\frac{\alpha\lambda^{2}}{s_{\ast}^{2}}>0.\end{equation*} This constraint is fulfilled only for negative $\alpha<0$. Therefore $P_6$ and $P_7$ are both saddle points. The critical point $P_3$ is the late-time attractor (see fig. \ref{fig02'}).

\subsubsection{$V=\frac{V_{0}}{(\eta + e^{-\alpha\phi})^{\beta}}$}

This potential (first studied in \cite{Zhou}) drives the evolution of the universe to transit from a scaling attractor into a de Sitter-like attractor. Following the above explained methodology we have:

\be f(s)=\frac{1}{\beta}+\frac{\alpha}{s},\label{f}\ee 

\be s_{\ast}=-\alpha\beta,\;df=-\frac{\alpha}{s_{\ast}^{2}}.\label{s}\ee

The condition $s_{\ast}df>0$ is satisfied whenever $\beta>0$, so that the possible late-time attractor solutions are the following:

\begin{itemize}

\item If $-\sqrt{\frac{3\gamma}{\beta^{2}}}<\alpha<\sqrt{\frac{6}{\beta^{2}}}$ the matter-scaling solution $P_7$ is a stable node.

\item If $\alpha>\sqrt\frac{6}{\beta^{2}}$ the matter-scaling solution $P_7$ is a stable spiral.

\end{itemize}

As in the above examples, from fig.\ref{fig02'} we see that the past attractor in the phase space is the empty, Misner-RS universe.

\begin{table*}\caption[crit]{Properties of the critical points for the autonomous system (\ref{eqx1}-\ref{eqs}).}
\begin{tabular}{@{\hspace{4pt}}c@{\hspace{14pt}}c@{\hspace{14pt}}c@{\hspace{14pt}}c@{\hspace{14pt}}c@{\hspace{14pt}}c@{\hspace{14pt}}c@{\hspace{14pt}}c@{\hspace{14pt}}c}
\hline\hline\\[-0.3cm]
$P_i$ &$x$&$y$&$z$&$s$&Existence& $\bar{\Omega}_\phi$& $\omega_\phi$& $q$\\[0.1cm]
\hline\\[-0.2cm]
$P_{1}$& $0$&$0$&$1$&$0$&both branches& $0$&undefined& $\frac{3\gamma-2}{2}$\\[0.2cm]
$P_{2}^\pm$& $\pm1$&$0$&$1$&$0$& " & $1$ & $1$ & $2$\\[0.2cm]
$P_{3}$& $0$&$1$&$z$&$0$& " & $1$ & $-1$ & $-1$\\[0.2cm]
$P_{4}$& $0$&$0$&$1$&$s_*$& always & $0$ &undefined& $\frac{3\gamma-2}{2}$\\[0.2cm]
$P_{5}^\pm$& $\pm1$&$0$&$1$&$s_*$& always & $1$ & $1$ & $2$ \\[0.2cm]
$P_{6}$& $\frac{s_*}{\sqrt{6}}$&$\sqrt{1-\frac{s_*^2}{6}}$&$1$&$s_*$& $s_*^2\leq 6$ & $1$ & $\frac{s_*^2-3}{3}$ & $\frac{s_*^2-2}{2}$\\[0.2cm]
$P_{7}$& $\frac{3}{2}\frac{\gamma}{s_*}$&$\sqrt{\frac{3\gamma(2-\gamma)}{2s_*^2}}$&$1$&$s_*$& always & $\frac{3\gamma}{s_*^2}$ & $-1+\gamma$ & $-1+\frac{3\gamma}{2}$\\[0.2cm]
\hline \hline
\end{tabular}\label{tab2}
\end{table*}
\begin{table*}\caption[eigenv]{Eigenvalues for the critical points in table \ref{tab2}($A\equiv\sqrt{(2-\gamma)[24\gamma^2-s_*^2(9\gamma-2)]}\;$).}
\begin{tabular}{@{\hspace{4pt}}c@{\hspace{14pt}}c@{\hspace{14pt}}c@{\hspace{14pt}}c@{\hspace{14pt}}c}
\hline\hline\\[-0.3cm]
$P_i$ & $\lambda_1$& $\lambda_2$& $\lambda_3$& $\lambda_4$\\[0.1cm]\hline\\[-0.2cm]
$P_{1}$& $0$ &$\frac{3}{2}(\gamma-2)$& $3\gamma/2$& $3\gamma/2$\\[0.2cm]
$P_{2}^\pm$& $3$ &$3$& $0$& $3(2-\gamma)$\\[0.2cm]
$P_{3}$& $-3$ & $0$ & $0$& $-3\gamma$\\[0.2cm]
$P_{4}$& $0$ &$\frac{3}{2}(\gamma-2)$& $3\gamma/2$& $3\gamma/2$\\[0.2cm]
$P_{5}^\pm$& $3$ & $\mp\sqrt{6}s_*^2\;df$ & $3(2-\gamma)$& $3\mp\frac{3}{2}s_*$\\[0.2cm]
$P_{6}$& $\frac{s_*^2-6}{2}$ & $-s_*^3\;df$ & $s_*^2-3\gamma$& $\frac{s_*^2}{2}$\\[0.2cm]
$P_{7}$& $-3s_*\;df$ & $\frac{3}{4s_*}[s_*(\gamma-2)-A]$ & $\frac{3}{4s_*}[s_*(\gamma-2)+A]$& $3\gamma/2$\\[0.2cm]
\hline \hline
\end{tabular}\label{tab3}
\end{table*}

\subsection{The Dvali-Gabadadze-Porrati braneworld}

We study here the stability of the critical points of the autonomous system (\ref{eqx1}-\ref{eqs}). As before, we will rely here on the assumption that the function $f(s)$ has zero(s) at given value(s) $s_*$: $f(s_*)=0$.

The critical points of the system (\ref{eqx1}-\ref{eqs}) are shown in table \ref{tab2}, while the eigenvalues of the corresponding Jacobian matrices are shown in the table \ref{tab3}. We have not included the points with $z=-1$ in our analysis, since the equations (\ref{eqx1}-\ref{eqs}) are invariant under the change of sign $z\rightarrow -z$.

The critical points $P_1,P_2^\pm,P_3$ are associated with the stationary points of the potential -- extrema and saddle stationary points -- and, in general, with potentials whose $\phi$-derivative vanishes (including the constant potential). The existence of the remaining critical points depends, in general, on the concrete functional form of the potentials, since, as stated before, the value $s_*$ is determined by the form of $f(s)$.

The points $P_1,P_2^\pm,P_3,P_4$ are non-hyperbolic critical points (one of the eigenvalues of the corresponding Jacobian matrixes vanishes). In this case the only thing we can state with certainty, on the basis of the straightforward analisys of the autonomous system of equations (\ref{eqx1}-\ref{eqs}) is that, depending on the phase considered -- the Minkowski phase or the self-accelerating one --, and on the initial conditions, trajectories in phase space originating in one of the repeller points ($P_5^\pm$), will inevitably approach one or several of the above non-hyperbolic critical points. 

Points $P_1$ and $P_4$ represent the matter-dominated solution, while $P_2^\pm$ and $P_5^\pm$ are associated with the solution dominated by the kinetic energy of the scalar field (the stiff-matter solution).\footnote{In fact, the points $P_1$ and $P_2^\pm$ are particular cases of $P_4$ and of $P_5^\pm$, respectively, when $s_*=0$.} They are linked always with decelerated expansion ($q=2$). The de Sitter-DGP -- accelerated -- solution corresponds to the critical point $P_3$. The point $P_6$, which exists whenever $s_*^2\leq 6$, is associated with the scalar field-dominated phase, while $P_7$ represents, in the phase space, the matter-scaling solution. 

Maximum we can say about the non-hyperbolic points is that they have attached an unstable subspace that is spanned by the eigenvectors: 

\bea v_1=\left(\begin{array}{clrrr}
0\\
0\\1\\0\end{array}\right),\;\;v_2=\left(\begin{array}{clrrr}
0\\
1\\0\\0 \end{array}\right),\;\;v_3=\left(\begin{array}{clrrr}
1\\
0\\0\\0 \end{array}\right).\nonumber \eea The point $P_5^-$ is always a past attractor in phase space, while $P_5^+$ is a saddle in $\Psi$. This is the classical result within general relativity with a minimally-coupled (self-interacting) scalar field.

The scalar field-dominated solution (point $P_6$) and the matter-scaling solution (critical point $P_7$), are always saddle points in the phase space. This result has to be contrasted with the classical result within general relativity with a minimally-coupled scalar field, where, depending on the values of the constant parameters, the above mentioned solutions can be late-time attractors in the phase space.

\begin{figure}[t!]
\begin{center}
\includegraphics[width=6.5cm,height=6cm]{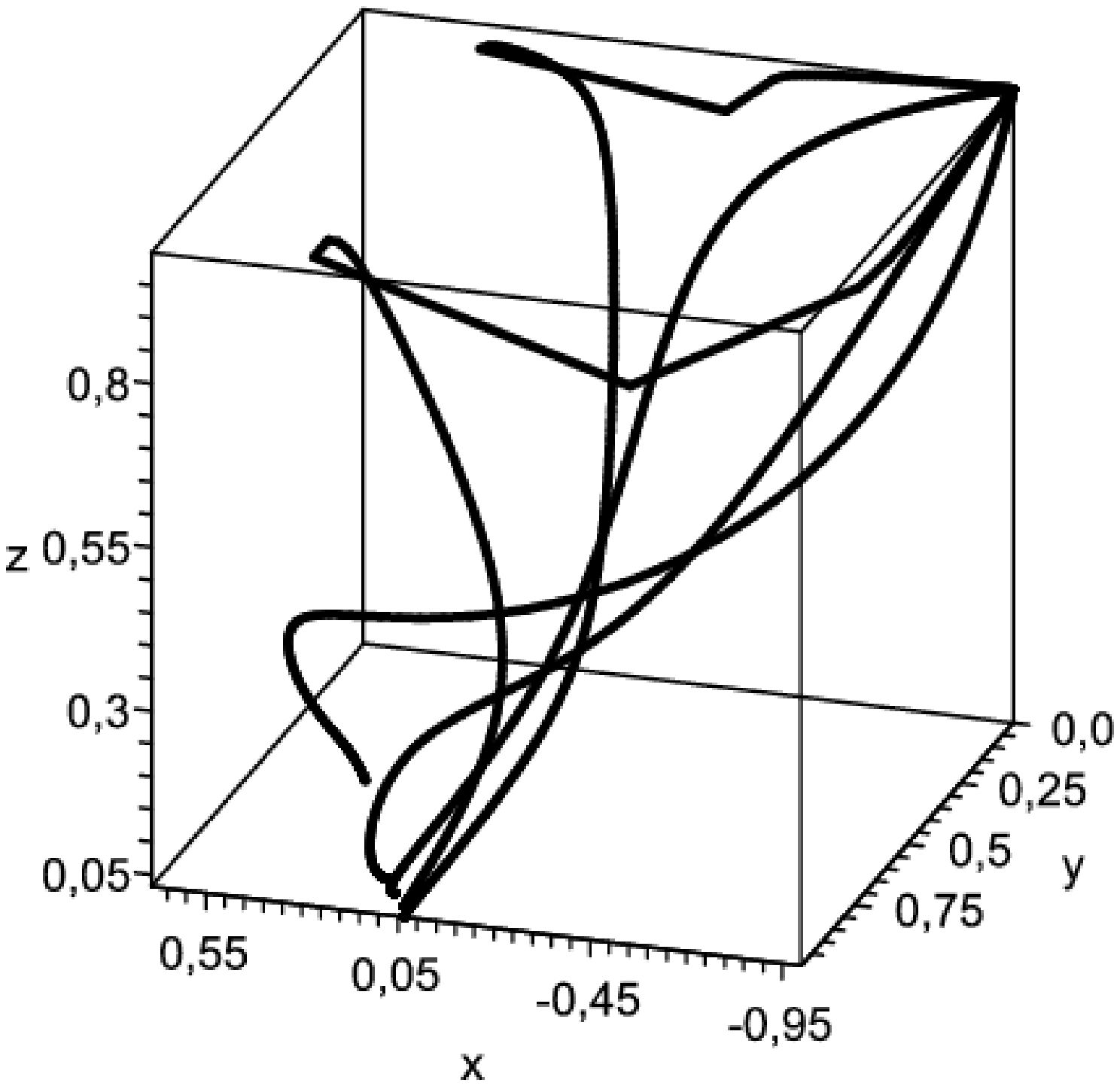}
\includegraphics[width=6.5cm,height=6cm]{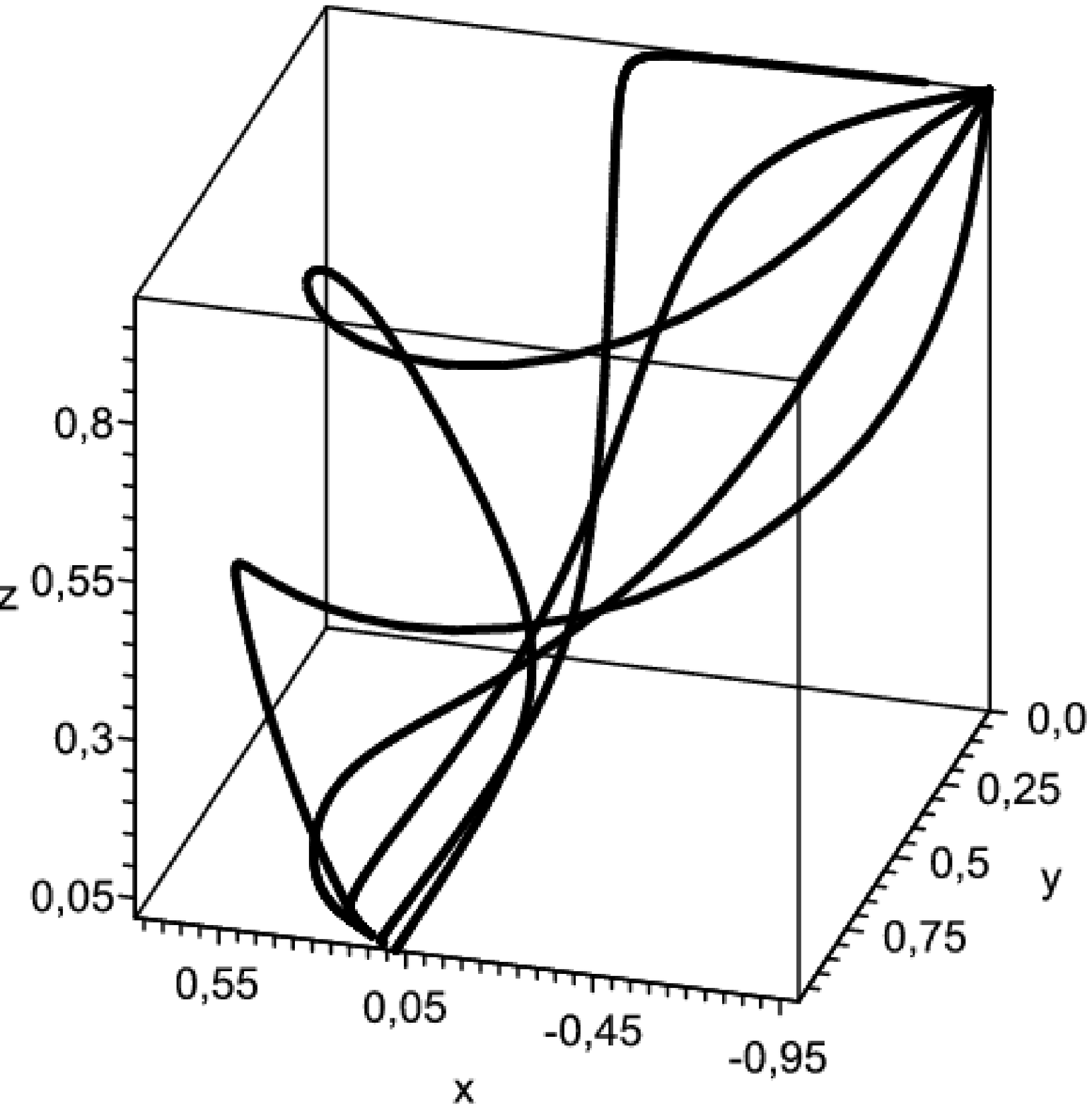}
\includegraphics[width=6.5cm,height=6cm]{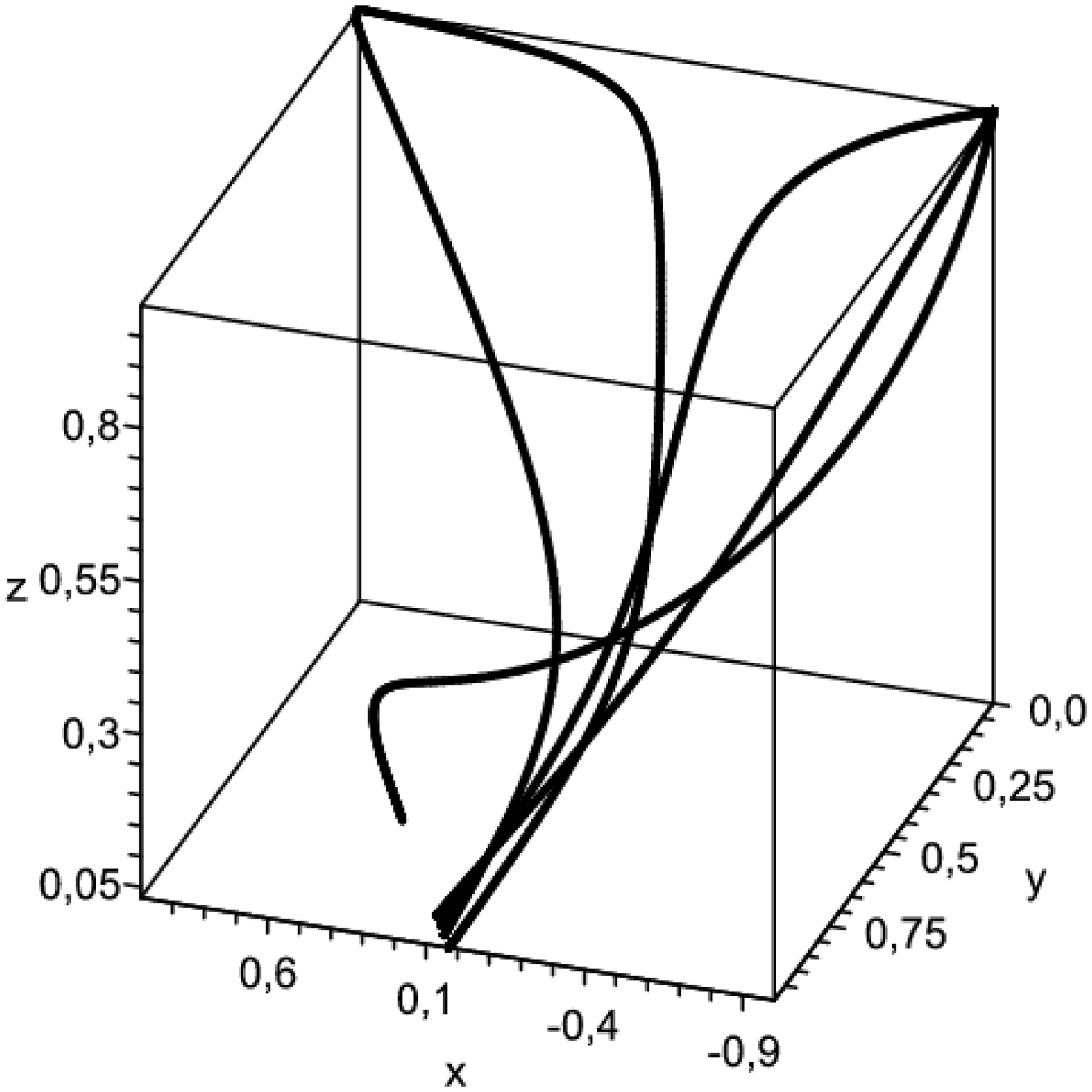}
\vspace{0.3cm}
\caption{Trajectories in phase space $(x,y,z)$ for different sets of initial conditions for the self-accelerating phase of the DGP model $\Psi_-$. The upper panel is for the potential $V=V_0[\sinh(\lambda\phi)]^{-\alpha}$, the panel at the center is for the potential $V=V_0\exp(\lambda\phi^2)/\phi^\alpha$, while the panel at the bottom is for the potential $V=V_0[\cosh(\lambda\phi)-1]^p$.}\label{fig01}
\end{center}
\end{figure}

The above mentioned results are illustrated in the figures \ref{fig01} and \ref{fig03}) for several quintessential potentials.

\begin{figure}[t!]
\begin{center}
\includegraphics[width=6.5cm,height=6cm]{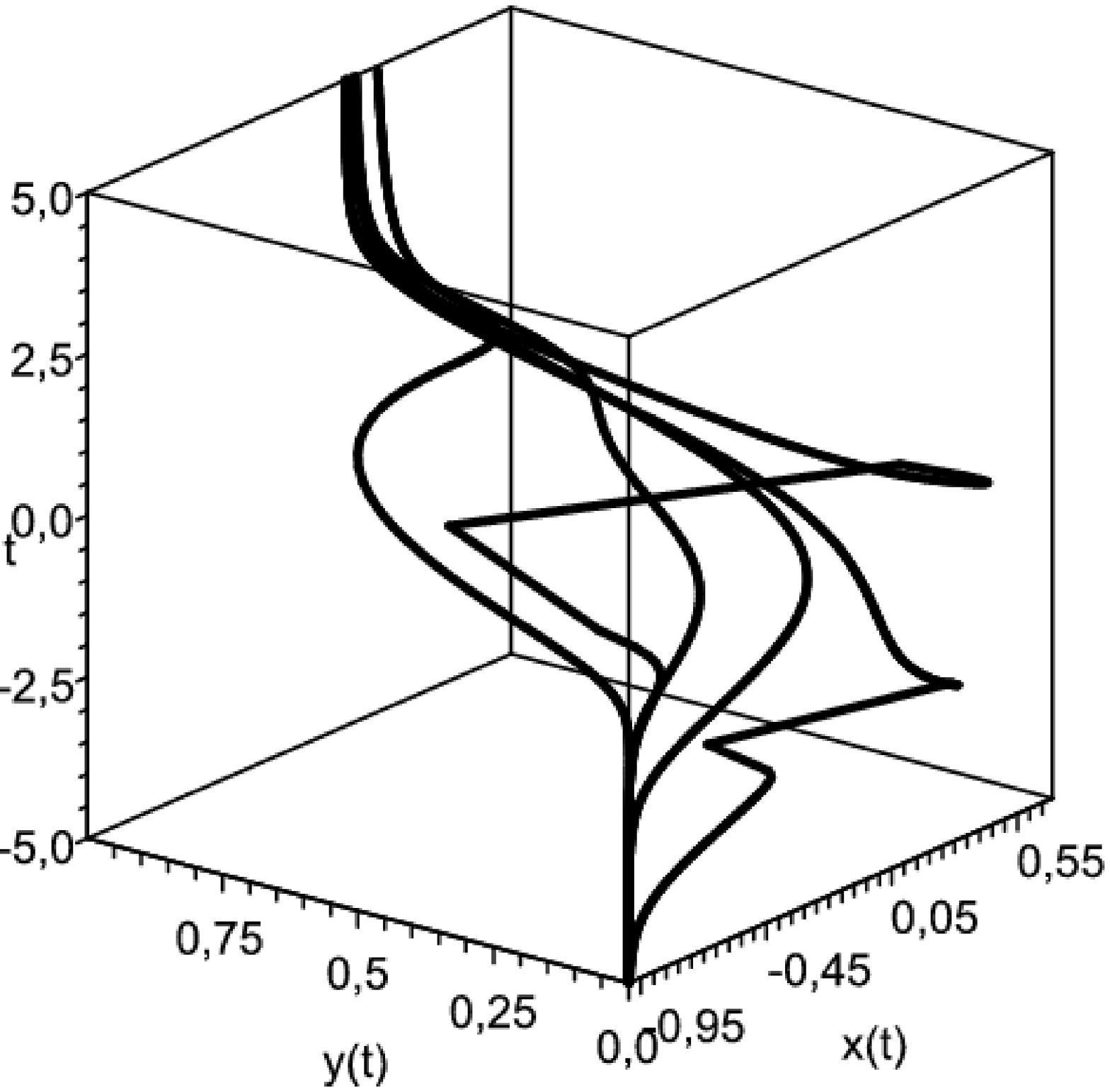}
\includegraphics[width=6.5cm,height=6cm]{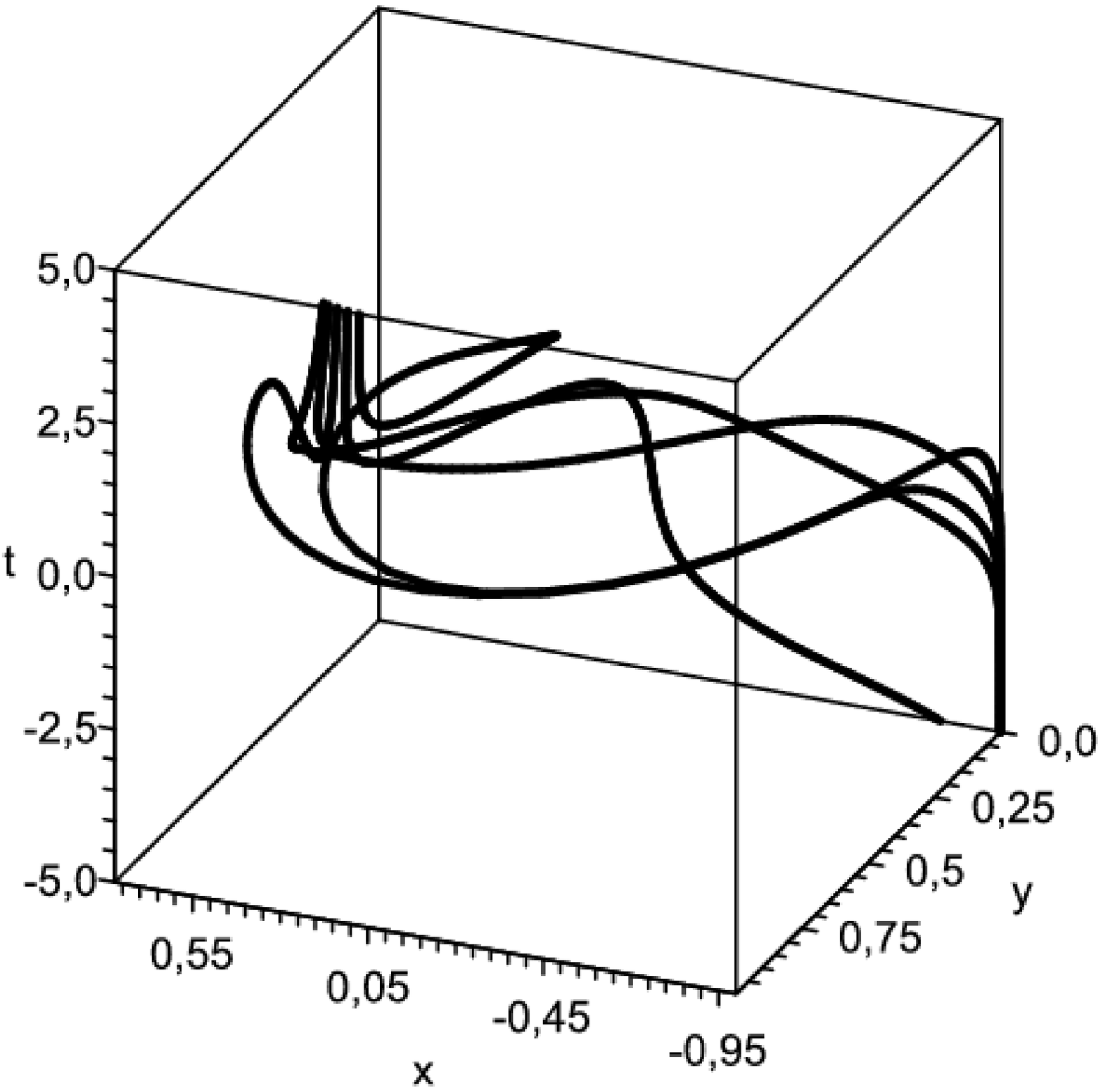}
\includegraphics[width=6.5cm,height=6cm]{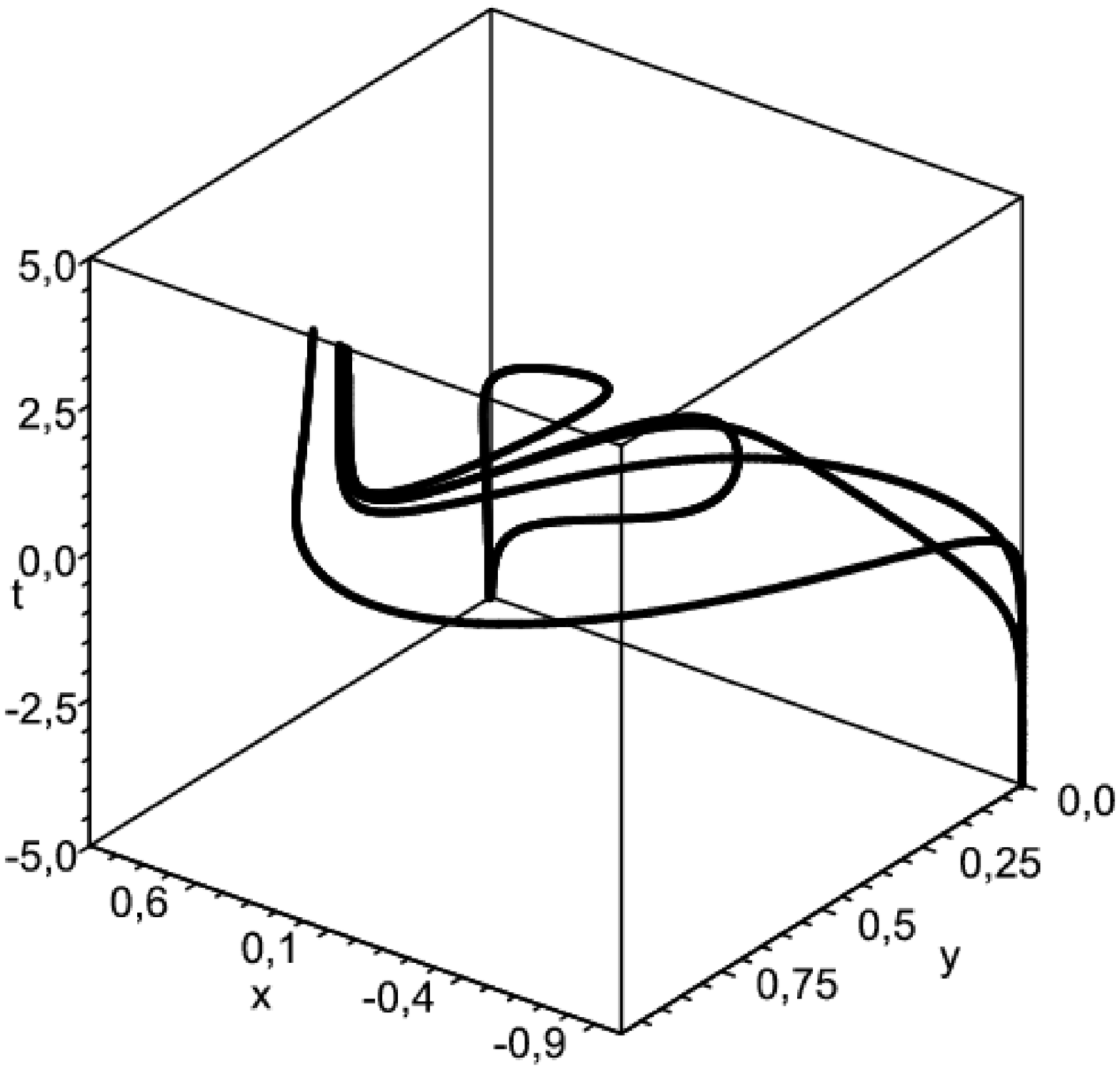}
\vspace{0.3cm}
\caption{Flux in time $(x(t),y(t),t)$ for the potentials in figure \ref{fig01}.}\label{fig02}
\end{center}
\end{figure}

\begin{figure}[t!]
\begin{center}
\includegraphics[width=6.5cm,height=6cm]{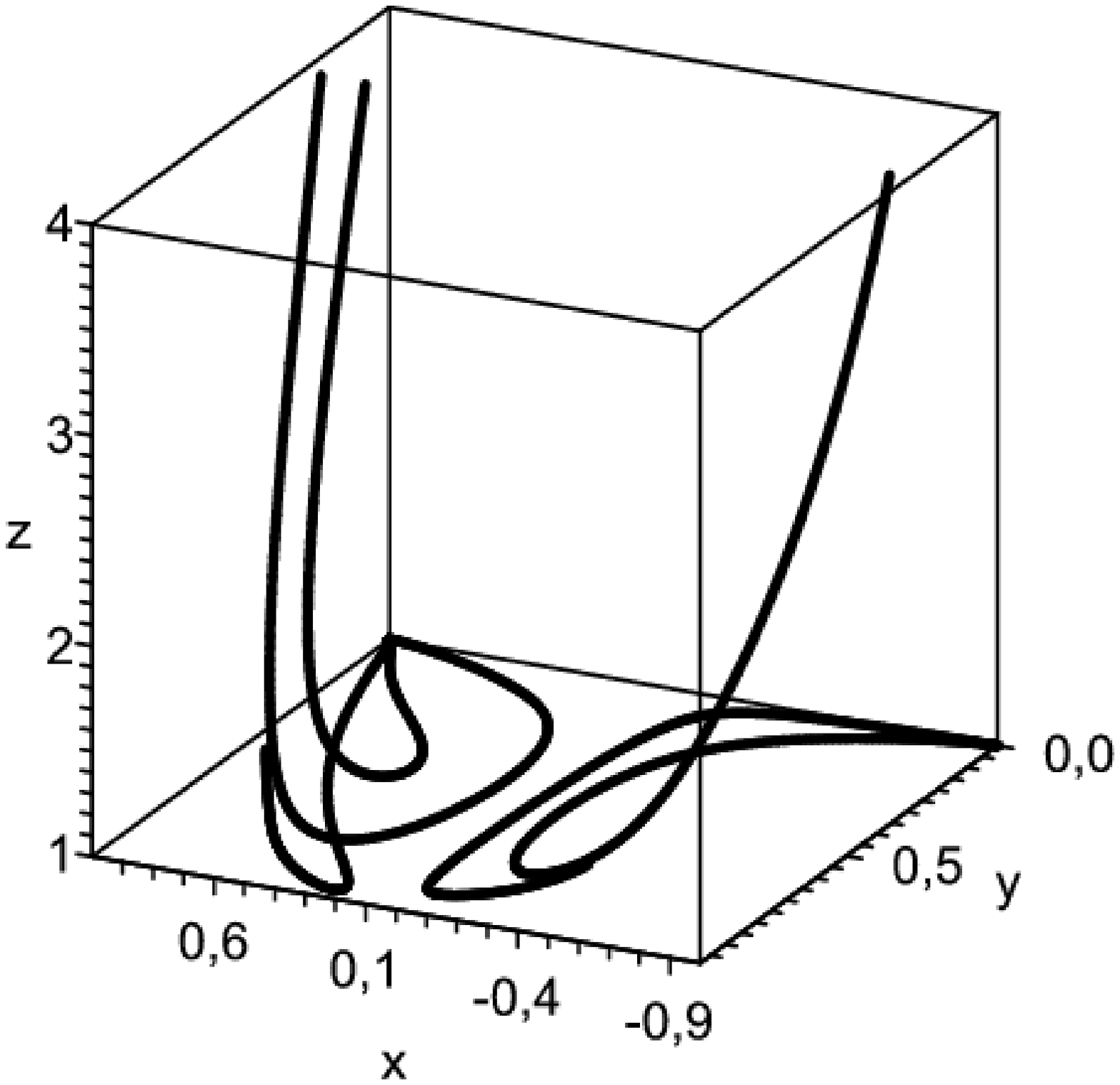}
\includegraphics[width=6.5cm,height=6cm]{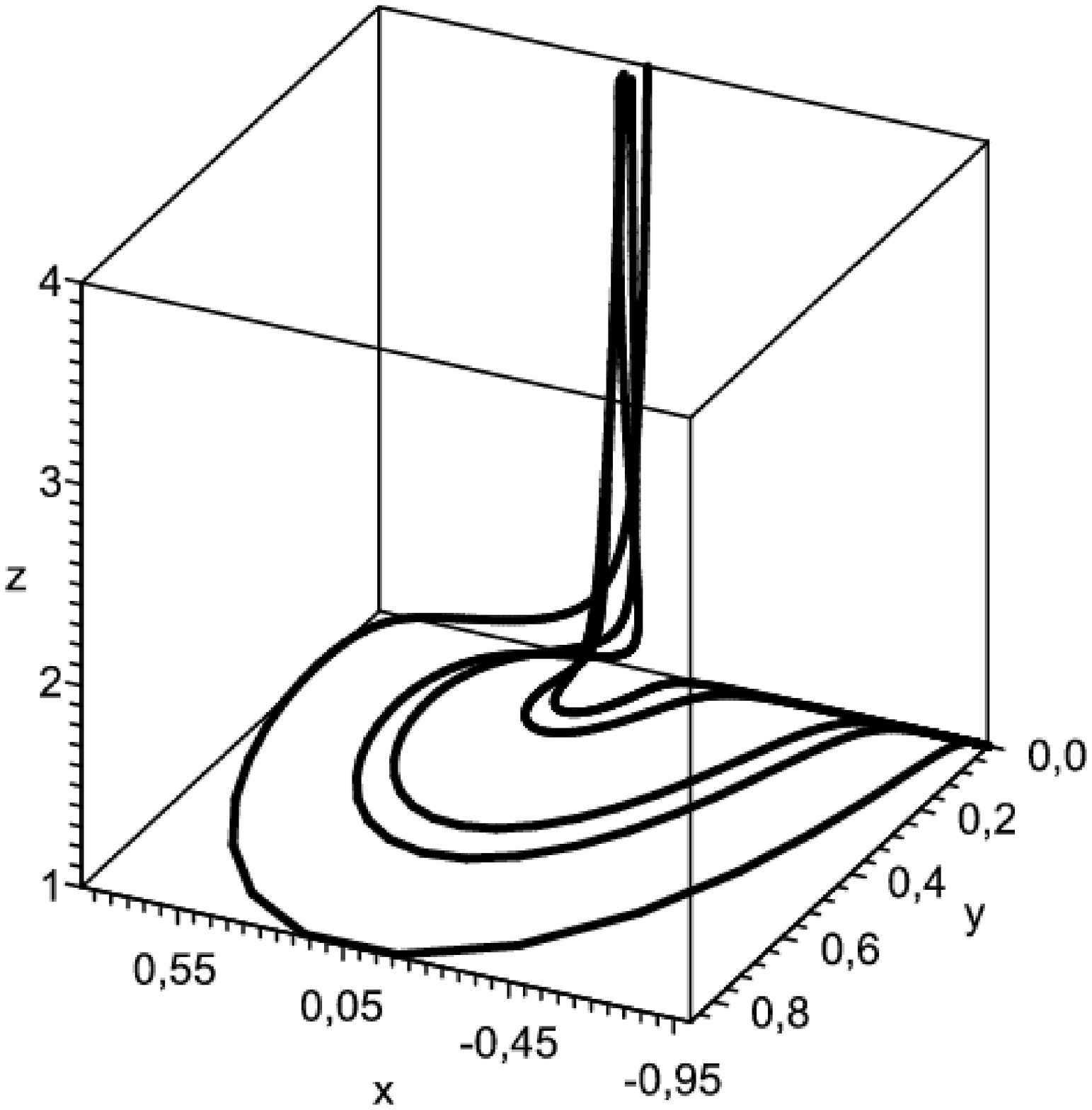}
\includegraphics[width=6.5cm,height=6cm]{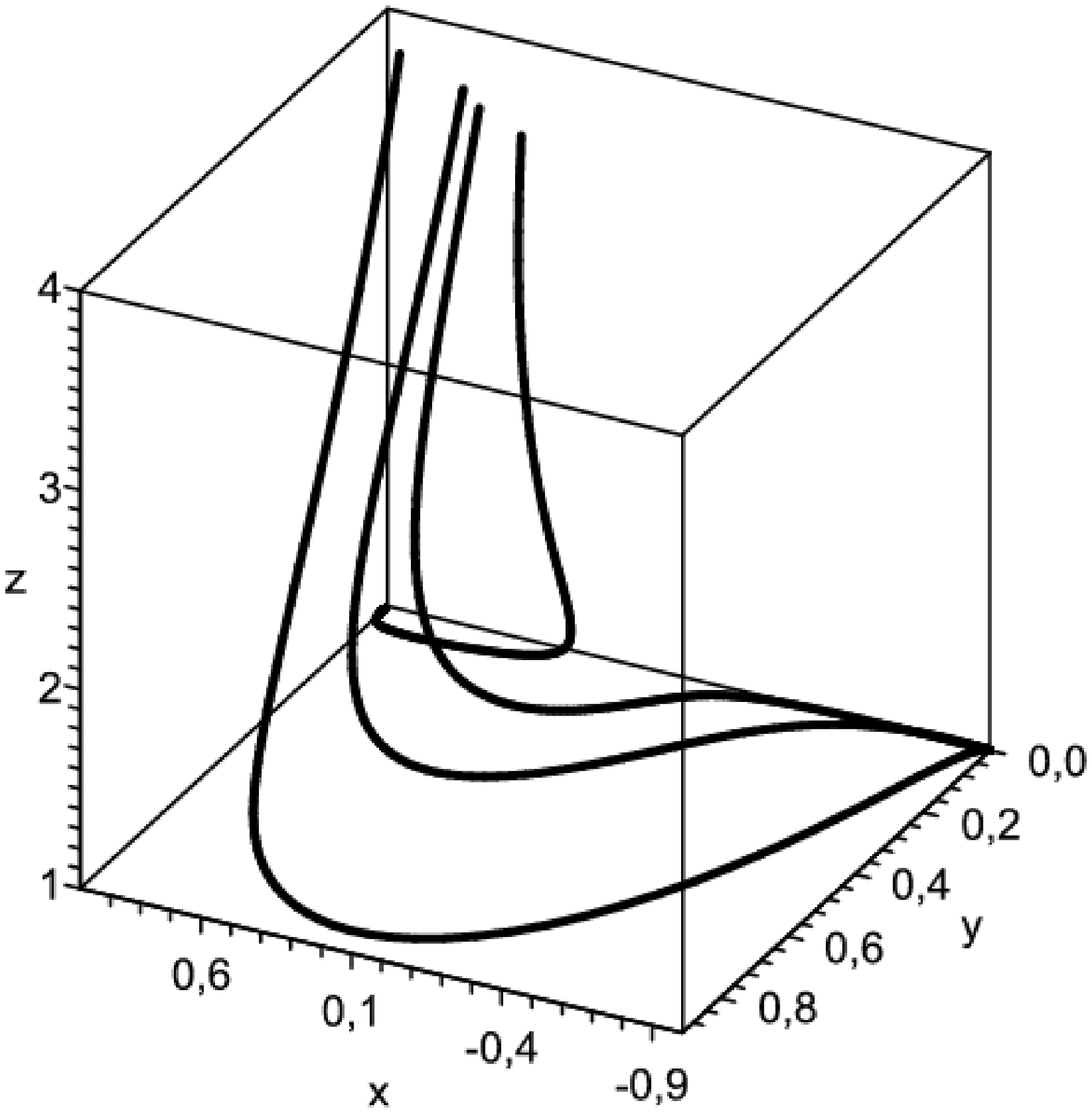}
\vspace{0.3cm}
\caption{Trajectories in phase space $(x,y,z)$ for different sets of initial conditions for the Minkowski phase of the DGP model $\Psi_+$. The upper panel is for the potential $V=V_0[\sinh(\lambda\phi)]^{-\alpha}$, the panel at the center is for the potential $V=V_0\exp(\lambda\phi^2)/\phi^\alpha$, while the panel at the bottom is for the potential $V=V_0[\cosh(\lambda\phi)-1]^p$ .}\label{fig03}
\end{center}
\end{figure}

For the potential $V=V_0[\sinh(\lambda\phi)]^{-\alpha}$ \cite{sahni1} (upper panel in the figure \ref{fig01}) $s_*=\pm\alpha\lambda$ and $df=\mp2/(\alpha^2\lambda)$, while for $V=V_0 \exp(\lambda\phi^2)/\phi^m$ \cite{brax} (a potential originated in supergravity models -- middle panel in fig. \ref{fig01}) $s_*=\pm\sqrt{-8m\lambda}$. The bottom panel is for the potential $V=V_0[\cosh(\lambda\phi)-1]^p$ \cite{sahni2}.\footnote{The author of reference \cite{sahni2} found that, for small values of $p$, ($p<1/2$), the scalar field dominates the mass density in the universe at late times, leading to accelerated expansion. This potential might serve as a good candidate for quintessence. The matter-scaling is approximately constant during the prolonged epoch.} In this case one have that, at $f(s)=0$, $s_*=\pm\lambda p$.

Phase trajectories in $\Psi_-$ (figure \ref{fig01}) originate from the source critical points $P_5^\pm$, corresponding to the standard 4D kinetic energy dominated (stiff-matter) solution and (asymptotically) approach to the point $(0,1,0,0)$ that has been removed from the phase space since phase space variables $x$ and $y$ blow up at the phase plane ($x,y,0,s$). The dynamics in the neigbourhood of this point has to be investigated in terms of different phase space variables.

Phase trajectories in $\Psi_+$ (figure \ref{fig03}) originate from the 4D stiff-matter solution (unstable node $P_5^\pm$ in tab. \ref{tab3}) and end up at the inflationary points $(0,1,z_{0i},0)\in P_3^\pm$, where the different $z_{0i}$-s are associated with the different initial conditions. Otherwise, points in $P_3^\pm$ are seen as attractor points by the different phase space "observers", moving along different phase trajectories that originate at $P_5^+$ or $P_5^-$.

\section{Results and discussion}

\subsection{RS model}

The main results of section IV.A can be summarized as follows:

\begin{itemize}

\item The matter-dominated solution (point $P_1$) is a non-hyperbolic critical point independent of the functional form of the self-interaction potential. It can be, at most, a saddle.

\item For $s_{\ast}^{2}<3\gamma$ ($s_{\ast}df>0$), the scalar field-dominated solution (point $P_6$) is the future atractor. Otherwise, $P_6$ is a saddle point. As seen from table \ref{tab1'} this critical point can be associated with accelerated expansion whenever $s_{\ast}^{2}< 2\gamma$ ($s_{\ast}df>0$).

\item For values $s_*^2>3\gamma$ the matter-scaling solution (point $P_7$) is a the late-time attractor. For $3\gamma<s_{\ast}^{2}<6$ it is a stable node, while, for $s_{\ast}^{2}>6$ it is a stable spiral. This solution is always a decelerating one.

\item The kinetic energy-dominated/stiff fluid solution (point $P_5^\pm$) is always a saddle critical point in the phase space.

\end{itemize}

The non-hyperbolic critical point $P_2$ (in fact, a set of critical points), represents the slow-roll Friedmann equation relating the Hubble expansion parameter with the potential of the inflaton field, modified by the presence of the RS brane (see equation (\ref{fr})). 

In general, the dynamical behavior of the Randall-Sundrum model differs from the standard behavior within four-dimensional Einstein-Hilbert gravity coupled to a self-interacting scalar field, only at early times (high energy regime). Actually, the empty (Misner-RS) universe is always the past attractor in the phase space of the Randall-Sundrum cosmological model \cite{isra}. This result is to be contrasted with the standard four-dimensional result where the kinetic energy-dominated solution is the past attractor \cite{wands,Zhou}. Within the present scenario the latter solution (critical points $P_4$ and $P_5^\pm$ in tables \ref{tab1'} and \ref{tab2'}) is always a saddle point.

The late-time cosmological dynamics, on the contrary, is not affected by the RS brane effects in any essential way.

\subsection{DGP model}

>From the analysis in section IV. B, the following important results can be summarized:

\begin{itemize}

\item Points $P_1-P_4$ in table \ref{tab2} are non-hyperbolic critical points so that, only the behaviour of the phase space trajectories may uncover the main properties of the dynamical system in their neighbourhood.

\item The kinetic energy-dominated solution can be either a past attractor (point $P_5^-$) or a saddle point in the phase space as in \cite{wands} -- see figure \ref{fig02}, where the time evolution of the dynamical system is shown -- since, at early times, the DGP brane effects can be safely ignored so that the standard cosmological dynamics is not modified. Recall that the DGP brane effects produce infra-red modifications to the laws of gravity.

\item The scalar field-dominated solution (critical point $P_6$ in Tab. \ref{tab2}), as well as the matter-scaling solution (point $P_7$), represent always saddle points in the phase space, contrary to the classical result within four-dimensional general relativity plus a minimally-coupled scalar field.

\end{itemize}

Apart from the exponential potential, there are a large number of potentials that can produce the matter-scaling solution (critical point $P_7$). This result is expected since, in the 4-dimensional limit, when standard Friedmann behavior is recovered, we are left with the case studied in the reference \cite{wands}, where the matter-scaling solution was identified as a critical point in phase space. 

Nevertheless, the DGP brane effects indeed modify the late-time cosmological dynamics through changing the stability of the corresponding (late-time) critical points. Actually, in the present case the matter-scaling solution (critical point $P_7$), as well as the scalar field-dominated phase, are always saddle critical points. This result has to be confronted with the classical general relativity result where the above mentioned solutions can be a late-time attractors.

\section{Conclusions}

In the present paper a thorough study of the phase space of both, the Randall-Sundrum and the Dvali-Gabadadze-Porrati braneworlds -- with a self-interacting scalar field trapped on the brane --, has been undertaken. A wide class of self-interaction potentials for which the quantity $\Gamma\equiv V\partial_\phi^2 V/(\partial_\phi V)^2$ can be written as a function of the variable $s\equiv -\partial_\phi V/V$, are included in this study.

It has been demonstrated, in particular, that the empty Misner-RS universe is always the past attractor in the phase space of the Randall-Sundrum cosmological model. The RS brane effects modify the early-time dynamics, so that, additionally, the kinetic energy-dominated solution -- the past attractor within general relativity plus a slef-interacting (minimally-coupled) scalar field -- is always a saddle critical point. The critical points that can be associated with late-time behaviour, as well as their stability properties, are not modified by the RS brane effects.

A detailed study of the dynamics of the DGP brane (with a self-interacting scalar field trapped on it), reveals that the critical points in phase space, coincide with the ones found in standard (four-dimensional) general relativity. An additional critical point that can be associated with five-dimensional behaviour, can be found only for the constant self-interaction potential. Nevertheless, even if, in general, there are no critical points that could be associated with genuine higher-dimensional effects, DGP brane effects indeed play a role: they modify the stability properties of the critical points associated with late-time cosmological dynamics.

The above results have been clearly illustrated with the help of phase space pictures generated by several potentials of cosmological interest.

This work was partially
supported by CONACyT M\'exico, under grants 49865-F and by grant number I0101/131/07 C-234/07, Instituto
Avanzado de Cosmologia (IAC) collaboration. Y L, D G, T G and I Q want to aknowledge the MES of Cuba by partial financial support of the present research.

\end{document}